\documentclass[aps,10pt,twocolumn,english,tightenlinesletterpaper,superscriptaddress,secnumarabic,nofootinbib]{revtex4}

\usepackage{amsmath}	
\usepackage{bm}		
\usepackage{amssymb}	
\usepackage{color} 		
\usepackage{graphicx} 	
\usepackage{times} 		
\usepackage{tensor}

\usepackage{hyperref} 	
\usepackage[all]{hypcap}  
\hypersetup{
    colorlinks=true,		
    linkcolor=black,		
    citecolor=blue,		
    filecolor=magenta,	
    urlcolor=black	
}


\newcommand{\eref}[1]{Eq.~(\ref{#1})}

\newcommand{\fref}[1]{Fig.~\ref{#1}}
\newcommand{\sref}[1]{Sec.~\ref{#1}}

\newcommand{\rref}[1]{Ref.~\cite{#1}}

\newcommand{\GeV}{\ \mathrm{GeV}}
\newcommand{\per}{\ . \ }
\newcommand{\com}{\ , \ }

\newcommand{\ie}{{\it i.e.}}
\newcommand{\eg}{{\it e.g.}}

\newcommand{\atCL}[1]{\ \text{at #1\% C.L.}}
\newcommand{\Lcal}{\mathcal{L}}
\newcommand{\Ncal}{\mathcal{N}}

\newcommand{\nn}{\nonumber \\}

\begin{document}

\title{Is Higgs Inflation Dead?}
\date{\today}

\author{Jessica L. Cook}
\affiliation{\small Department of Physics and School of Earth and Space Exploration \\ Arizona State University, Tempe, AZ 85827-1404}

\author{Lawrence M. Krauss}
\email{krauss@asu.edu}
\affiliation{\small Department of Physics and School of Earth and Space Exploration \\ Arizona State University, Tempe, AZ 85827-1404}
\affiliation{\small Research School of Astronomy and Astrophysics, Mt. Stromlo Observatory, \\ Australian National University, Canberra, Australia 2611}

\author{Andrew J. Long}
\email{andrewjlong@asu.edu}
\affiliation{\small Department of Physics and School of Earth and Space Exploration \\ Arizona State University, Tempe, AZ 85827-1404}

\author{Subir Sabharwal}
\affiliation{\small Department of Physics and School of Earth and Space Exploration \\ Arizona State University, Tempe, AZ 85827-1404}

\begin{abstract}
We consider the status of Higgs Inflation in light of the recently announced detection of B-modes in the polarization of the cosmic microwave background radiation by the BICEP2 collaboration.  
In order for the primordial B-mode signal to be observable by BICEP2, the energy scale of inflation must be high, $V_{\rm inf} \approx 2 \times 10^{16} \GeV$.  Higgs Inflation generally predicts a small amplitude of tensor perturbations, and therefore it is natural to ask if Higgs Inflation might accommodate this new measurement.   We find the answer is essentially no, unless one considers either extreme fine tuning, or possibly adding new beyond the standard model fields, which remove some of the more attractive features of the original idea.  We also explore the possible importance of a factor that has not previously been explicitly incorporated, namely the gauge dependence of the effective potential used in calculating inflationary observables, e.g. $n_S$ and $r$, to see if this might provide additional wiggle room.  
Such gauge effects are comparable to effects of Higgs mass uncertainties and other observables already considered in the analysis, and therefore they are relevant for constraining models.  
But, they are therefore too small to remove the apparent incompatibility between the BICEP2 observation and the predictions of Higgs Inflation.

\end{abstract}

\maketitle

\section{Introduction}\label{sec:Introduction}

The theory of inflation \cite{Starobinsky:1980te, Guth:1980zm, Linde:1981mu} successfully addressed the twentieth century's greatest puzzles of theoretical cosmology.  
Over the past 20 years, increasingly precise measurements of the temperature fluctuations of the cosmic microwave background radiation (CMB) also confirmed the nearly scale invariant power spectrum of scalar perturbations, a relatively generic inflationary prediction.  
These many successes, however, underscored the inability to probe perhaps the most robust and unambiguous prediction of inflation, the generation of a background of gravity waves associated with what are likely enormous energy densities concomitant with inflation ({\it e.g.},  \cite{Krauss:1992ke}).  

Recently, the BICEP2 collaboration reported evidence of B-modes in the polarization pattern of the CMB \cite{Collaboration:2014fk}.  
The B-modes result from primordial gravity wave induced distortions at the surface of last scattering.  
If one assumes that these gravity waves are of an inflationary origin, then the BICEP2 measurement corresponds to an energy scale of inflation:  
\begin{align}\label{eq:BICEP}
	V_{\rm inf}^{1/4} \approx (2\pm 0.2) \times 10^{16} \GeV
\end{align}
for a reported tensor-to-scalar ratio of $r \approx 0.2 ^{+ 0.07} _{-0.05}$ (using also the Planck collaboration's measurement of the amplitude of the scalar power spectrum \cite{Ade:2013rta}).  
Such a high scale of inflation rules out many compelling models. 
For the purposes of this paper, we will assume that the observation $r \approx 0.2$ is valid\footnote{Note that the BICEP2 measurement is in tension with the upper bound, $r < 0.11 \atCL{95}$, obtained previously by the Planck collaboration \cite{Ade:2013rta}.  }, and we assess the impact of this measurement on a particular model of inflation, known as Higgs Inflation.  

Higgs Inflation ({\rm HI}) postulates that the Standard Model Higgs field and the inflaton are one in the same \cite{Bezrukov:2007ep}.  
(See also \rref{Bezrukov:2013fka} for a recent review).  
This powerful assumption allows {\rm HI} to be, in principle much more predictive than many other models of inflation, as by measuring the masses of the Higgs boson and the top quark at the electroweak scale ($100 \GeV$), one might predict observables at much larger 
energy scales 
associated with inflation 
($V_{\rm inf}^{1/4} \lesssim 10^{16} \GeV$).  

However, in practice this enhanced predictive power is elusive due to a strong sensitivity to quantum effects, unknown physics, and other technical subtleties in the model.  
Specifically, one connects observables at the electroweak and inflationary scales using the renormalization group flow (RG) of the SM couplings \cite{DeSimone:2008ei, Barvinsky:2008ia, Bezrukov:2010jz, Bezrukov:2012sa, Allison:2013uaa, Salvio:2013rja}.  
It is reasonable however to expect that there is new physics at intermediate scales, and even if the SM is extended only minimally to include a dark matter candidate \cite{Clark:2009dc} or neutrino masses \cite{Okada:2009wz, He:2010sk, Rodejohann:2012px, Kobakhidze:2013pya} this new physics can qualitatively affect the connection between electroweak and inflationary observables.  
Moreover, perturbative unitarity arguments require new physics just above the scale of inflation \cite{Barbon:2009ya, Burgess:2010zq}, and in addition the unknown coefficients of dimension six operators can significantly limit the predictive power of {\rm HI} \cite{Burgess:2014lza}.  
The {\rm HI} calculation also runs into various technical subtleties that arise from the requisite non-minimal gravitational coupling (see below) and quantization in a curved spacetime \cite{George:2012xs,George:2013iia, Prokopec:2014iya}.  
Finally, it is worth noting that {\rm HI} is also at tension with the measured Higgs boson and top quark masses, and an $O(2\sigma)$ heavier Higgs or lighter top is required to evade vacuum stability problems \cite{Buttazzo:2013uya}.  

Also, as we shall later discuss in detail, there is one additional source of ambiguity in calculations of {\rm HI} that had not been fully explored.  
Since the quantum corrections are significant when connecting the low energy and high energy observables, one should not work with the classical (tree-level) scalar potential, as is done in may models of inflation, but one must calculate the quantum effective potential.  
It is well-known that in a gauge theory the effective potential explicitly depends upon the choice of gauge in which the calculation is performed \cite{Weinberg:1973ua, Jackiw:1974cv}, and care must be taken to extract gauge-invariant observables from it \cite{Dolan:1974gu, Kang:1974yj, Nielsen:1975fs, Fukuda:1975di} (see also \cite{Aitchison:1983ns, Patel:2011th}).  
This fact can perhaps be understood most directly by recalling that the effective action is the generating functional for one-particle irreducible Green's functions, which themselves are gauge dependent \cite{Jackiw:1974cv}.  
In practice one often neglects this subtlety, fixes the gauge at the start of the calculation, and calculates observables with the effective potential as if it were a classical potential.  
In the context of finite temperature phase transitions, it is known that when calculated naively in this way, the predictions for observables depend on the choice of gauge used \cite{Weinberg:1974hy, Dolan:1973qd, Bernard:1974bq, Patel:2011th, Wainwright:2011qy, Wainwright:2012zn, Garny:2012cg}.  
Because of the extreme tension between {\rm HI} models and the data, we assess here the degree to which this gauge uncertainty might affect the observables in Higgs Inflation.  
We find that the gauge ambiguity introduces uncertainties that are comparable to the variation of the physical parameters, {\it i.e.} the Higgs mass.  
As a result, this ambiguity alone cannot resuscitate moribund models.

\section{Gravity Waves from Higgs Inflation}\label{sec:GravityWaves}

The Standard Model Higgs potential, $V(h) = \lambda h^4 / 4$ with $\lambda=O(0.1)$, is too steeply sloped for successful inflation.  
The measurement of the Higgs boson mass fixes $\lambda \approx 0.13$, whereas $\lambda \ll 1$ is required to produce the observed amplitude of density perturbations.  
In the {\rm HI} model, slow roll is achieved by introducing a non-minimal gravitational coupling for the Higgs field, $\Lcal = - \xi \, \Phi^{\dagger} \Phi R$, where $\Phi$ is the Higgs doublet and $R$ the Ricci scalar.  
One can remove the non-minimal coupling term from the Lagrangian by performing a conformal transformation, $g_{\mu \nu}(x) = \Omega^{-2}(x) \hat{g}_{\mu \nu}(x)$ where
\begin{align}\label{eq:Omega_def}
	\Omega^2 = 1 + 2 \xi \Phi^{\dagger} \Phi / M_P^2 
\end{align}
is the conformal factor and $M_P$ is the reduced Planck mass.  
By doing so, one passes from the Jordan to the Einstein frame.  
The scalar potential in the new frame becomes
\begin{align}\label{eq:HI_potential}
	V(h) = \frac{\lambda h^4}{4 \left( 1 + \frac{\xi h^2}{M_P^2} \right)^2}
\end{align}
where we have written $\Phi^{\dagger} \Phi = h^2 / 2$.  
At large field values, $h \gg M_P / \sqrt{\xi}$, the potential asymptotes to a constant
\begin{align}\label{eq:V0}
	V_0 \approx \lambda M_P^4 / 4 \xi^2 \per
\end{align}
This is the appropriate regime for slow roll inflation.  

To evince the tension between Higgs Inflation and large tensor perturbations we can first neglect quantum corrections to $V(h)$, {\it e.g.} the running of $\lambda$, as the energy scale of {\rm HI}, given by \eref{eq:V0}, is insensitive to the quantum corrections, whereas the slope is more sensitive.  

Since $\lambda$ is fixed by the measured Higgs mass, the scalar potential in \eref{eq:HI_potential} has only one free parameter: $\xi$.  
It is well-known that to achieve sufficient e-foldings of inflation and the correct amplitude for the scalar power spectrum, one needs the non-minimal coupling to be much larger than unity.  
Specifically one requires (see, \eg, \rref{Bezrukov:2013fka}) 
\begin{align}
	\xi \approx 47000 \sqrt{\lambda} 
\end{align}
which is $\xi \approx 17000$ for $\lambda \approx 0.13$.  
The energy scale of inflation is then predicted to be
\begin{align}
	V_0 \approx (0.79 \times 10^{16} \GeV)^4
\end{align}
leading to a tensor-to-scalar ratio, assume scalar density perturbations fixed by CMB observations, 
$r \approx 0.0036$.  
This is naively incompatible with the much larger BICEP2 measurement, see \eref{eq:BICEP}.  
Decrease $\xi$ in {\rm HI} to attempt to match the newly measured value of $V_{\rm inf}$ is not workable either, as 
setting $\xi \approx 2000$ then produces too little power in scalar density perturbations.  

Fundamentally then, the problem in obtaining a large value of $r$ in Higgs inflationary models is that the {\rm HI} potential asymptotes to a constant at large field values where inflation occurs.  This flat potential then results in relatively large density perturbations, which, in order to then match observations, constrain the magnitude of the potential, resulting in a small tensor contribution.

The question then becomes whether variations in this canonical {\rm HI}, due to quantum effects for example, will allow the SM Higgs boson to the be inflaton field while also accommodating the large value of $r$.

\section{Saving Higgs Inflation?}\label{sec:Saving}

Since it is the non-minimal coupling, $\xi$, that flattens out the potential at high scales, one might consider whether there are other ways to flatten the potential, and so avoid the requirement for large $\xi$ values.

One possibility proposed in this regard ~\cite{Allison:2013uaa} involves fine tuning the Higgs and top masses such that the Higgs self-coupling runs very small at the scale of inflation, $\lambda \sim 10^{-4}$.  
This allows for relatively small $\xi \sim 90$ and produces $r \gtrsim 0.15$ that may be compatible with the BICEP measurement.  
It is impossible to entirely eliminate the need for the non-minimal coupling.  
However, as a caveat let us point out that this solution only exists if the theory is first quantized in the Jordan frame and then moved to the Einstein frame (so-called ``prescription I''), and results differ if the operations are reversed (``prescription II'').  
The apparent disagreement is an artifact of quantizing all the fields except gravity, which results in a different definition of the Ricci scalar in the two frames.  
A full theory of quantum gravity would probably be required to resolve the problem consistently between frames.  
Thus, it is not clear if the small $\xi$ ``prescription I'' solution is artificial.  

If one goes outside of the Standard Model, then new physics can affect the running of the Higgs self-coupling or anomalous dimension, $\gamma$.  
For example, one may hope that $\lambda$ or $\gamma$ acquires a significant running at high scales so as to give a workable solution consistent with both the measured scalar and tensor power spectra.  
(See, {\it e.g.}, \cite{Joergensen:2014rya}).  

As a result, it appears that canonical {\rm HI} with a non-minimal gravitational coupling as the only new physical input appears extremely difficult to reconcile with the new observation of a large tensor contribution from inflation. 
If would appear to be necessary to add new physics to eliminate the dependence on non-minimal coupling entirely and to give the Higgs effective potential a shape compatible with observations.  
Such extension of {\rm HI} tend to defeat the original purpose of the idea, namely its predictivity, and in any case most such modifications that have been proposed \cite{Lebedev:2011aq, Khoze:2013uia, Hamada:2013mya} tend to retain the now undesirable feature of small $r$ in any case. 

There are two options that might allow large $r$ consistent with BICEP.  
One possibility involves tuning the Higgs potential to form a second local minimum at large scales, {\it i.e.}, a false vacuum similar to old inflation ~\cite{Masina:2011un}.  
To avoid the problems of old inflation, a time dependent tunneling rate is introduced.  
While most mechanisms to achieve this, however, produce a small value of $r$ ~\cite{Masina:2012yd}, large $r$ can be accommodated by adding a new scalar with a non-minimal coupling to gravity, such that the Higgs field sees a time dependent Planck mass~\cite{Masina:2011aa}.  
A second possibility uses a non-canonical Galileon type kinetic term for the Higgs field.  
This model yields an $r \simeq 0.14$~\cite{Kamada:2010qe}. 

These tuned limits, variants, and extensions of the original {\rm HI} model leave the door slightly open for the possibility of connecting the Higgs with the inflation field.  
However, without additional scalars or modification of the Higgs potential via some other mechanism beyond the Standard Model, the original scenario, {\it i.e.} Higgs Inflation with only a non-minimal coupling to gravity, does not appear to be compatible with the BICEP result. 

Before we nail the coffin shut on Higgs Inflation, however, there is one possible additional source of uncertainty that merits further investigation. As we describe below, when one goes beyond the tree level, there are gauge ambiguities involved in the calculation of effective potentials that need to be considered when deriving constraints on parameters.  

\section{Gauge Dependence Ambiguities}\label{sec:Gauge}

When working with a gauge theory, such as the Standard Model electroweak sector, calculations typically involve spurious gauge dependence that cancels when physical observable are calculated.  
For example, in a spontaneously broken Yang-Mills theory one may work in the renormalizable class of gauges ($R_{\xi}$) upon augmenting the Lagrangian with a gauge fixing term $\Lcal_{gf} = - G^a G^a /2$ where $G^a = (1/\sqrt{\xi_{\rm gf}})(\partial_{\mu} A^{a \, \mu} - \xi g \tensor{F}{^{a}_{i}} \chi_{i})$ where $\chi_{i}$ are the would-be Goldstone boson fields and $\tensor{F}{^{a}_{i}} = T^{a}_{ij} v_j$ with $T^{a}_{ij}$ the symmetry generators and $v_{j}$ the symmetry-breaking vacuum expectation value.  
(See, \eg, \cite{Aitchison:1983ns}).  
A corresponding Fadeev-Popov ghost term is also added.  
Physical or ``on-shell'' quantities, such as cross sections and decay rates, may be calculated perturbatively, and any dependence on the gauge fixing parameter, $\xi_{\rm gf}$, cancels order-by-order.  
Unphysical or ``off-shell'' quantities, such as propagators or one-particle irreducible Green's functions, may harmlessly retain the spurious gauge dependence.  

The Coleman-Weinberg effective action $\Gamma_{\rm eff}$ and effective potential $V_{\rm eff}$ \cite{Coleman:1973jx} have become standard tools in the study of vacuum structure, phase transitions, and inflation.  
The effective action is the generating functional of one-particle irreducible Green's functions, and therefore it is important to recognize that both $\Gamma_{\rm eff}$ and $V_{\rm eff}$ are off-shell quantities, which will carry spurious gauge dependence \cite{Jackiw:1974cv}.  
When applying the effective potential to a problem, special care must be taken to extract gauge-invariant information.  
In particular, the Nielsen identities express the gauge invariance of the effective potential at its stationary points, but derivatives of the effective potential are not generally gauge invariant \cite{Nielsen:1975fs}.  
This suggests that inflationary observables, e.g. $n_S$, $r$, and $dn_S / d\ln k$, naively extracted directly from the slow roll parameters will acquire a spurious gauge dependence.  

Ideally one would like to determine the ``correct'' procedure for calculating physical quantities like $n_S$ from a given model in such a way that the spurious gauge dependence is canceled.  
There have been significant efforts made in this direction \cite{George:2012xs,George:2013iia}, but a full gauge invariant formalism is yet to be developed.  
Here we will take a different approach that is more aligned with recent work on the gauge dependence of phase transition calculations \cite{Patel:2011th, Wainwright:2011qy, Wainwright:2012zn}.  
Specifically, we numerically perform the ``naive'' {\rm HI} calculation using the $R_{\xi}$ gauge effective potential and RG-improvement to assess the sensitivity of the inflationary observables to the spurious gauge dependence.  

We begin by reviewing the familiar Higgs Inflation calculation.  
After moving from the Jordan to the Einstein frame, as described in \sref{sec:GravityWaves}, the resulting action contains a non-canonical kinetic term for the Higgs field.  
One cannot, in general, find a field redefinition that makes the kinetic term canonical globally \cite{Kaiser:2010ps, Burgess:2010zq}.  
At this point, it is customary to move to the unitary gauge where the Higgs doublet is written as $\Phi(x) = e^{2 i \pi^a(x) \tau^a} (0 \, , \, h(x) / \sqrt{2} )^T$.  
Then the kinetic term for the radial Higgs excitation can be normalized by the field redefinition $\chi(h)$ where 
\begin{align}
	d\chi/dh = \sqrt{ \frac{1}{\Omega^2} + \frac{3}{2} \frac{ M_P^2 (d\Omega^2 / dh)^{2} }{\Omega^2} }
\end{align}
and now $\Omega^2 = 1 + \xi h^2 / M_P^2$.  

Having canonically normalized both the gravity and Higgs kinetic terms, the derivation of the effective potential proceeds along the standard lines.  
We calculate the RG improved, one-loop effective potential as described in the Appendix.  
After performing the RG improvement, the parameter $\lambda$ that appears in \eref{eq:HI_potential} should be understood at the running coupling evaluated at the scale of inflation.  
Generally, $\lambda < 0.1$ and its value depends upon the physical Higgs boson and top quark masses at the input scale.  
For the best fit observed values, $M_H \approx 125 \GeV$ and $M_t \approx 173 \GeV$, the coupling runs negative at $h \approx 10^{10}-10^{12} \GeV$; this is the well-known vacuum stability problem of the Standard Model \cite{Buttazzo:2013uya}.  
Successful {\rm HI} requires an $O(2\sigma)$ deviation from central values toward either larger Higgs boson mass or smaller top quark mass.  

Gauge dependence enters the calculation at two places:  explicitly in the one-loop correction to the effective potential and implicitly through the Higgs anomalous dimension upon performing the RG improvement.  

To calculate the slow roll parameters, {\it e.g.}
\begin{align}\label{eq:slow_roll}
	&\epsilon = \frac{M_P^2}{2} \bigl( V^{\prime} / V \bigr)^2 \bigr|_{h_{\rm cmb}} 
\end{align}
the derivatives are taken with respect to $\chi$, \ie, $V^{\prime}(h(\chi)) = (\partial V / \partial h) (d\chi / dh)^{-1}$.  
The potential and its derivatives are evaluated at the field value, $h_{\rm cmb}$, for which the number of e-foldings, given by 
\begin{align}\label{eq:N_efold}
	\mathcal{N} = \int_{h_{\rm end}}^{h_{\rm cmb}} d h \, \frac{V(h)}{V^{\prime}(h) M_P^2} \com
\end{align}
is $\Ncal = 60$.  
Inflation terminates at $h = h_{\rm end}$ where $(M_P^2 / 2) ( V^{\prime} / V )^2 = 1$.  

\begin{figure}[h]
\begin{center}
\includegraphics[width=0.45\textwidth]{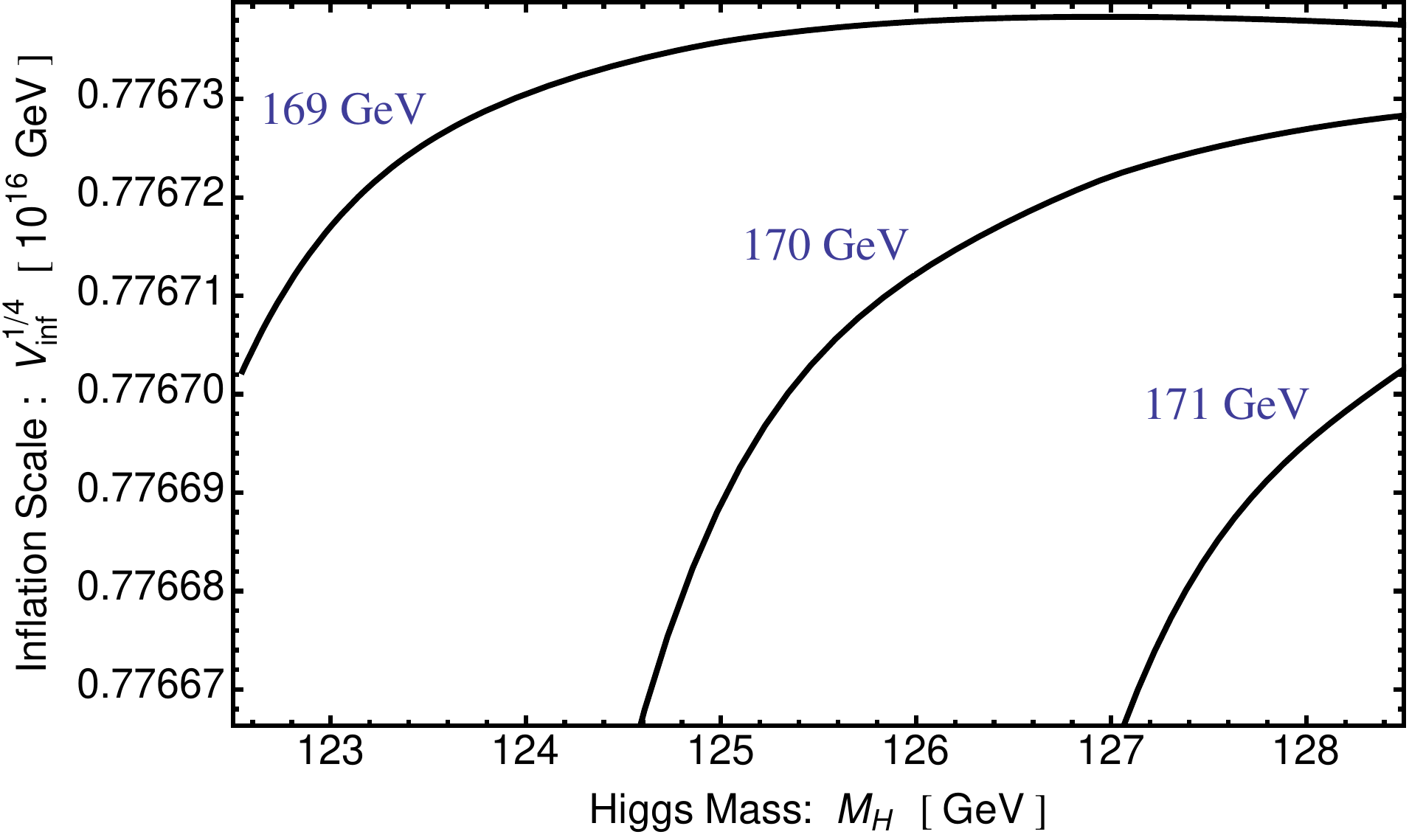} 
\caption{\label{fig:plot_r}
The predicted energy scale of inflation, $V_{\rm inf}^{1/4}$, over a range of Higgs boson masses ($M_H$), for three values of the top quark mass ($M_t$), and in the Landau gauge, $\xi_{\rm gf}=0$.  
}
\end{center}
\end{figure}

In \fref{fig:plot_r} we show the energy scale of inflation, 
\begin{align}
	V_{\rm inf} = V(h_{\rm cmb}) \com
\end{align}
as the the Higgs boson and top quark masses are varied, and the non-minimal coupling, $\xi \approx {\rm few} \times 10^{3}$, is determined to match the observed amplitude of scalar perturbations.  
This demonstrates that the scale of inflation is insensitive to $M_H$, varying only at the $O(10^{-4})$ level.  
It always remains significantly below $2 \times 10^{16} \GeV$, which indicates the incompatibility with the BICEP2 measurement.  
(The corresponding tensor-to-scalar ratio is $r \approx 0.003$.)  

\begin{figure}[h]
\begin{center}
\includegraphics[width=0.45\textwidth]{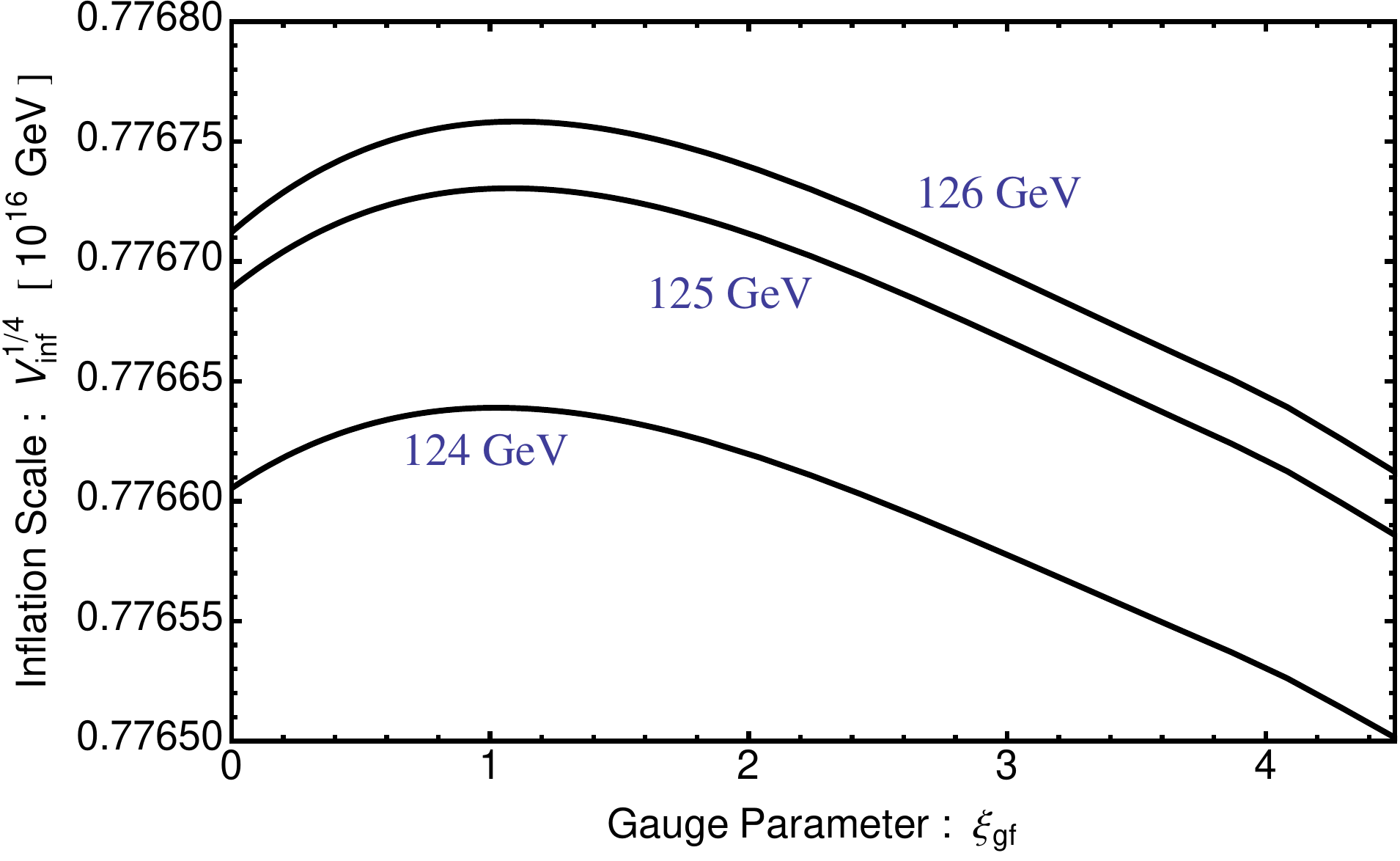} 
\caption{\label{fig:vary_xi}
The energy scale of inflation, $V_{\rm inf}$, as the gauge parameter, $\xi_{\rm gf}$, varies.  We fix $M_t = 170 \GeV$ and show three values of $M_H$.  
}
\end{center}
\end{figure}

To illustrate the gauge dependence, we show in \fref{fig:vary_xi} how $V_{\rm inf}$ varies with $\xi_{\rm gf}$.  
We find that $V_{\rm inf}$ also changes at a level comparable to its sensitivity to $M_H$ or $M_t$ as the gauge parameter deviates from the Landau gauge ($\xi_{\rm gf} = 0$).  
It is therefore important to consider this ambiguity for model building purposes.  
Nevertheless, the absolute change in $V_{\rm inf}$ is far too small to reconcile {\rm HI} with the BICEP2 measurement.  

Note that at larger vales of $\xi_{\rm gf}$ the scale of inflation appears to continue to decrease, but in this limit the perturbative validity of the calculation begins to break down.  
To resolve this issue, the unphysical degrees of freedom, the Goldstone bosons and ghosts, should be decoupled as the unitary gauge is approached.  

Our numerical results appear consistent with the Nielsen identities \cite{Nielsen:1975fs, Fukuda:1975di} which capture the gauge dependence of the effective potential.  
The relevant identity is
\begin{align}\label{eq:Nielsen_iden}
	\left[ \xi \frac{\partial}{\partial \xi} + C(\phi, \xi) \frac{\partial}{\partial \phi} \right] V_{\rm eff}(\phi,\xi) = 0 \per
\end{align}
In the slow roll regime, the gradient of the effective potential is small, and the gauge dependence is proportionally suppressed.  

We note that a rigorous gauge invariant calculation could perhaps take \eref{eq:Nielsen_iden} as a starting point.  
This might be an interesting avenue for future work, either in the context of {\rm HI} or other, potentially more viable models of inflation that are embedded in gauge theories.  

\section{Conclusion}\label{sec:Conclusion}

The recent detection of B-modes by the BICEP2 collaboration represents a profound and exciting leap forward in our ability to explore fundamental physics and the early universe. 
If the measurement of $r \approx 0.2$ is confirmed, then it is reasonable to expect that, in the not-too-distant future, measurements of the spectrum of primordial tensor perturbations will become possible, allowing further tests of inflation. And if the measured $r$ can unambiguously be shown to be due to inflation, then this also substantiates the quantization of gravity \cite{Krauss:2013pha}.  

Thus, future observations will provide significant constraints on particle physics and models of inflation. However the simple observation of non-zero $r$ already signals the death knell for low-scale models of inflation.  This includes the class of models captured by the potential in \eref{eq:HI_potential}, and among these apparently Higgs Inflation.  
We have shown that $r \approx 0.2$ essentially excludes canonical Higgs Inflation in the absence of extreme fine tuning.  The Higgs field may live on as the inflaton but only with significant non-minimal variants of {\rm HI}.  

In our analysis we have also drawn attention to the issue of gauge dependence in the Higgs Inflation calculation.  
We find that the energy scale of inflation acquires an artificial dependence on the gauge fixing parameter by virtue of the gauge dependence of the effective potential from it is extracted.  
However, we find this gauge dependence of the scale of inflation is comparable to the dependence on other physical parameter uncertainties, which are themselves small.  
While this may be important for model building purposes, it does not affect the robustness of the fact that large $r$ disfavors Higgs Inflation.

\acknowledgments
This work was supported by ANU and by the US DOE under Grant No.\  DE-SC0008016. We would like to thank Jayden L. Newstead for help with the code.  

\appendix

\section{Standard Model Effective Potential}\label{app:SM_Veff}

The Standard Model effective potential is calculated (i) to the one-loop order, (ii) working in the $\overline{\rm MS}$ renormalization scheme with renormalization scale $\mu$, and (iii) in the renormalizable class of gauges ($R_{\xi}$) as follows:  
\begin{align}\label{eq:Veff_SM_V0}
	V_{\rm eff}(h) = V^{(0)}(h) + V^{(1)}(h) \per
\end{align}
The tree-level potential is 
\begin{align}
	V^{(0)}(h) & = \frac{\lambda}{4} h^4 \com
\end{align}
and we can neglect the $O(h^0)$ and $O(h^2)$ terms for the purposes of studying {\rm HI} where the field value is large.  
The one-loop correction is \cite{Sher:1988mj} (see also \cite{Patel:2011th} for gauge dependent factors)
\begin{align}\label{eq:Veff_SM_V1}
	&V^{(1)}(h) = 
	-\frac{12}{4} \frac{\tilde{m}_t^4}{16 \pi^2} \left( \ln \frac{\tilde{m}_t^2}{\mu^2} - \frac{3}{2} \right) \\
	& \qquad
	+ \frac{6}{4} \frac{\tilde{m}_W^4}{16 \pi^2} \left( \ln \frac{\tilde{m}_W^2}{\mu^2} - \frac{5}{6} \right)
	+ \frac{3}{4} \frac{\tilde{m}_Z^4}{16 \pi^2} \left( \ln \frac{\tilde{m}_Z^2}{\mu^2} - \frac{5}{6} \right) \nn
	& \qquad 
	+ \frac{1}{4} \frac{\tilde{m}_G^4}{16 \pi^2} \left( \ln \frac{\tilde{m}_G^2}{\mu^2} - \frac{3}{2} \right) 
	+ \frac{2}{4} \frac{\tilde{m}_{G^{\pm}}^4}{16 \pi^2} \left( \ln \frac{\tilde{m}_{G^{\pm}}^2}{\mu^2} - \frac{3}{2} \right) \nn
	& \qquad
	- \frac{2}{4} \frac{\tilde{m}_{c_W}^4}{16 \pi^2} \left( \ln \frac{\tilde{m}_{c_W}^2}{\mu^2} - \frac{3}{2} \right) 
	- \frac{1}{4} \frac{\tilde{m}_{c_Z}^4}{16 \pi^2} \left( \ln \frac{\tilde{m}_{c_Z}^2}{\mu^2} - \frac{3}{2} \right) \nonumber
\end{align}
where we have neglected the light fermions.  
We also neglect the contribution from the Higgs mass term. During inflation, the potential is very flat and this contribution is subdominant.  The remaining SM fields, the massless photon and gluons, do not enter the effective potential at the one-loop order.  
The effective masses are 
\begin{align}
\begin{array}{ccl}
	\text{ Top Quark} 
	& \quad & 
	\tilde{m}_{t}^2 = \frac{y_{t}^2}{2 \Omega^2} h^2 \\
	\text{ W-Bosons} 
	& \quad & 
	\tilde{m}_{W}^2 = \frac{g^2}{4 \Omega^2} h^2 \\
	\text{ Z-Bosons} 
	& \quad & 
	\tilde{m}_{Z}^2 = \frac{g^2 + g^{\prime \, 2}}{4 \Omega^2} h^2 \\
	\text{ Higgs Boson} 
	& \quad & 
	\tilde{m}_{H}^2 = \frac{3 \lambda}{\Omega^4} h^2 \frac{1 - \xi h^2 / M_P^2}{\Omega^2 + 6 \xi^2 h^2 / M_P^2} \\
	\text{ Neutral Goldstone} 
	& \quad & 
	\tilde{m}_{G}^2 = \frac{\lambda}{\Omega^4} h^2 + \tilde{m}_{c_Z}^2 \\
	\text{ Charged Goldstones} 
	& \quad & 
	\tilde{m}_{G^{\pm}}^2 = \frac{\lambda}{\Omega^4} h^2 + \tilde{m}_{c_W}^2 \\
	\text{ Ghosts} 
	& \quad & 
	\tilde{m}_{c_Z}^2 =\xi_{\rm gf} \tilde{m}_Z^2 \\
	\text{ Ghosts} 
	& \quad & 
	\tilde{m}_{c_W}^2 =\xi_{\rm gf} \tilde{m}_W^2 
\end{array} 
\end{align}
where $\Omega^2 = 1 + \xi h^2 / M_P^2$ was given by \eref{eq:Omega_def}.  
We denote the gauge fixing parameter by $\xi_{\rm gf}$ to distinguish it from the non-minimal gravitational coupling parameter, $\xi$.  

We implement the RG improvement as per \cite{Kastening:1991gv, Bando:1992np, Bando:1992wy}.  
(See also the reviews \cite{Sher:1988mj, Ford:1992mv}).  
This consists of (1) solving the RG equations (RGEs) to determine the running parameters as functions of the RG flow parameter $t$, (2) replacing the various coupling constants in $V_{\rm eff}$ with the corresponding running parameter, and (3) evaluating the RG flow parameter at the appropriate value $t=t_{\ast}$ so as to minimize the would-be large logarithms.  

For the sake of discussion, let us denote the running parameters collectively as $\hat{c}_i(t) = \bigl\{ \hat{g}_3(t), \hat{g}_2(t), \hat{g}_1(t) , \hat{\lambda}(t), \hat{y}_t(t), \hat{\xi}(t) \bigr\}$ where $g_{2} = g$ and $g_{1} = g^{\prime}$. Then the RGEs take the form $\beta_{\hat{c}_i}/(1+\gamma) = d \hat{c}_i / dt$ with the boundary condition $\hat{c}_i(t=0) = c_{i,0}$. Here $\gamma$ is the anomalous dimension of the Higgs field. 
We neglect the running of the gauge-fixing parameter, $\xi_{\rm gf}$, since it is self-renormalized.  
This approximation is reasonable since we focus on $\xi_{\rm gf} < 4\pi$; for larger values of $\xi_{\rm gf}$, perturbativity becomes an issue.  
The Higgs field runs according to $- \gamma \hat{h} = d \hat{h}/dt$ where the anomalous dimension $\gamma(t)$ is given as \cite{Bednyakov:2013eba} 
\begin{align}\label{eq:anom_dim}
	\gamma & = \ 
	\frac{1}{(4\pi)^2} \Bigl[ - \frac{9}{4} \Bigl( 1 - \frac{\xi_{\rm gf}}{3} \Bigr) g_2^2 - \frac{3}{4} \Bigl( 1 - \frac{\xi_{\rm gf}}{3} \Bigr) g_1^2 + 3 y_t^2 
	\Bigr] \nn 
	& - \frac{1}{(4\pi)^4} \Bigl[ \Bigl( \frac{271}{32} -3\xi_{\rm gf} - \frac{3}{8}{\xi^2_{\rm gf}} \Bigr) g_2^4- \frac{9}{16} g_1^2 g_2^2 -6\lambda^2 \nn
	&- \frac{431}{96} g_1^4-\frac{5}{2} (\frac{9}{4} g_2^2+ \frac{17}{12} g_1^2+8g_3^2)y_t^2+ \frac{27}{4} y_t^4 \Bigr] \per
\end{align} 
This last equation may be solved immediately along with the boundary condition $\hat{h}(t=0) = h_c$ to obtain 
\begin{align}
	\hat{h}(t) = h_c \, e^{\hat{\Gamma}(t)}
\end{align}
where $\hat{\Gamma}(t) = - \int_{0}^{t} \, \gamma(t^{\prime})/(1+\gamma(t^{\prime})) \, d t^{\prime}$, and we seek to calculate the effective potential as a function of $h_c$.  
The beta functions are independent of $\xi_{\rm gf}$, but the anomalous dimension is gauge-variant since the Higgs field is a gauge-variant operator.  
Finally, the renormalization scale runs according to $\hat{\mu} = d \hat{\mu} / dt$, which may be solved along with $\hat{\mu}(t=0) = \mu_0$ to obtain $\hat{\mu}(t) = \mu_0 e^t$.  

We solve the one-loop beta functions using the Mathematica code made publicly available by Fedor Bezrukov at \url{http://www.inr.ac.ru/~fedor/SM/} .  
The code implements the matching at the electroweak scale to determine the couplings, $c_{i,0}$, at the scale $\mu_0 = M_t$ in terms of the physical masses and parameters.  
The code was extended (1) by generalizing the anomalous dimension to the $R_{\xi}$ gauge as in \eref{eq:anom_dim}, and (2) by including the field-dependent factors of
\begin{align}\label{eq:comm_factor}
	s = \frac{1 + \frac{\hat{\xi}(t) \hat{h}(t)^2}{M_P^2}}{1 + (1 + 6 \hat{\xi}(t))  \frac{\hat{\xi}(t) \hat{h}(t)^2}{M_P^2}}
\end{align}
in the two-loop beta functions, as indicated by \cite{Allison:2013uaa}.  
The factor of $s$ arises because of the non-canonical Higgs kinetic term, and it appears in the commutator of the Higgs field with its conjugate momentum \cite{DeSimone:2008ei}.  

Finally the RG-improved effective potential is evaluated as in \eref{eq:Veff_SM_V0} after making the replacements $\lambda \to \hat{\lambda}(t_{\ast})$, $g \to \hat{g}(t_{\ast})$, $h \to \hat{h}(t_{\ast})$, $\mu \to \hat{\mu}(t_{\ast})$, and so on.  
The RG flow parameter, $t_{\ast}$, is chosen to minimize the would-be large logarithm arising from the top quark.  
This is accomplished by solving 
\begin{align}\label{eq:tast_def}
	\frac{\hat{y}_t(t)^2 \hat{h}(t)^2}{2 (1 + \frac{\hat{\xi}(t) \hat{h}(t)^2}{M_P^2} ) \hat{\mu}(t)^2} \Bigr|_{t=t_{\ast}} = 1 \com
\end{align}
which must be done numerically.  
Note that $t_{\ast}$ is an implicit function of the field variable, $h_c$.  
This can be seen by writing 
\begin{align}
	t_{\ast} = \frac{1}{2} \ln \left[ \frac{\hat{y}_t(t_{\ast})^2 e^{2 \hat{\Gamma}{(t_{\ast})}} h_c^2}{2 \mu_0^2} \right]
	\approx \frac{1}{2} \ln \left[ \frac{y_0^2 h_c^2}{2 \mu_0^2} \right] \per
\end{align}
Using \eref{eq:tast_def}, the commutator factor in \eref{eq:comm_factor} is written as 
\begin{align}
	s = \Bigl[ 1 + 12 \frac{\hat{\xi}(t)^2 \hat{\mu}(t)^2}{\hat{y}_t^2 M_P^2}  \Bigr]^{-1} \com
\end{align}
and the field dependence drops out.  

\bibliographystyle{h-physrev5}
\bibliography{refs--Higgs_Inflation}

\end{document}